\documentclass[10pt]{article}
\usepackage{fullpage,graphicx,subfigure,mathdots,mathpazo,color}
\usepackage{amsmath,amscd,tikz,mathrsfs,cite}
\usepackage[normalem]{ulem}
\usepackage{amsmath}
\usepackage{setspace}

\usepackage{hhline}  
\usepackage{epsfig,amsmath,graphicx,amssymb,overpic}
\usepackage{bm}
\usepackage{graphicx}
\usepackage{subfigure, xcolor}
\usepackage{ntheorem}
\usepackage{listings,diagbox}
\usepackage[title]{appendix} 
\usepackage{color}
\usepackage{epsfig,amsmath,graphicx,amssymb,overpic}

\usepackage{booktabs}
\usepackage{diagbox}
\usepackage{graphicx}
\usepackage{dcolumn}
\usepackage{bm}
\usepackage{graphicx}
\usepackage{subfigure}
\usepackage{epsfig,amsmath,graphicx,amssymb,overpic,cite}
\usepackage{makecell}

\def\be{\begin{equation}}
\def\ee{\end{equation}}
\def\bee{\begin{eqnarray}}
\def\ene{\end{eqnarray}}
\def\bes{\begin{subequations}}
\def\ees{\end{subequations}}

\allowdisplaybreaks[4]

\begin{document}

\baselineskip=14pt \renewcommand {\thefootnote}{\dag}
\renewcommand
{\thefootnote}{\ddag} \renewcommand {\thefootnote}{ }

\pagestyle{plain}


\begin{center}
\baselineskip=16pt \leftline{}
\baselineskip=16pt \leftline{} \vspace{-.3in} {\Large \textbf{{\ Asymmetric integrable turbulence and rogue wave statistics for \\ the derivative nonlinear Schr\"odinger equation}}} \\[0.2in]

Ming Zhong$^{1,2}$, Weifang Weng$^{3}$  and Zhenya Yan$^{\mathrm{*,4,2}}$
\\[0.05in]
{\small $^1$  School of Advanced Interdisciplinary Sciences, University of Chinese Academy of Sciences, Beijing 100049, China
 $^2$State Key Laboratory of Mathematical Sciences, Academy of Mathematics and Systems Science,\\ Chinese Academy of Sciences, Beijing 100190, China \newline
$^3$School of Mathematical Sciences, University of Electronic Science and Technology of China, Chengdu 611731, China\newline
$^4$ School of Mathematics and Information Science, Zhongyuan University of Technology, Zhengzhou 450007, China \footnote{$^{*}$Corresponding author.
\textit{Email address}: zyyan@mmrc.iss.ac.cn}}
\end{center}
{\baselineskip=13pt }


\vspace{0.2in}

\noindent
{\bf Abstract.\, }We investigate a comprehensive analysis of asymmetric integrable turbulence and rogue waves (RWs) emerging from the modulation instability (MI) of plane waves for the derivative nonlinear Schr\"odinger (DNLS)  equation. Over extended temporal evolution, a range of statistical measures are employed to assess the turbulence characteristics, including the \(n\)-th moments, ensemble-averaged  kinetic and potential energy, wave-action spectrum,  probability density function (PDF), and auto-correlation function.  The \(n\)-th moments and ensemble-averaged kinetic and potential energy exhibit oscillatory convergence towards their steady-state values. Specifically, the amplitudes of oscillations for these indexes  decay asymptotically with time as \(t^{-1.36}\), while the phase shifts demonstrate a nonlinear decay with a rate of \(t^{-0.78}\). The frequency of these oscillations is observed to be twice the maximum growth rate of  MI. These oscillations can be classified into two distinct types: one is in phase with ensemble-averaged potential energy modulus $|\langle H_4\rangle|$, and the other is anti-phase. At the same time, this unity is also reflected in the wave-action spectrum \( S_k(t) \) for a given \( k \), the auto-correlation function \( g(x,t) \) for a given \( x \), as well as the PDF \( P(I,t) \)  for a given \( I = |\psi|^2 \). The critical feature of the turbulence is the wave-action spectrum, which follows a power-law distribution of the form \( |k+3|^{-\alpha} \) expect for the finite value at $k=-3$. Unlike the NLS equation, the turbulence in the DNLS setting is asymmetric, primarily due to the asymmetry between the wave number of the plane wave from the MI and the perturbation wave number.. As the asymptotic peak value of \( S_k \) is observed at \( k = -3 \), the auto-correlation function exhibits a nonzero level as \( x \to \pm L/2 \). The PDF of the wave intensity asymptotically approaches the exponential distribution in an oscillatory manner. However, during the initial stage of the nonlinear phase, MI slightly increases the occurrence of RWs. This happens at the moments when the potential modulus is at its minimum, where the PDF becomes ``fatter", and the probability of RWs occurring in the range of \( I\in [12, 15] \) is significantly higher than in the asymptotic steady state.

\vspace{0.1in} \noindent Keywords: \thinspace \thinspace\ Integrable turbulence, Derivative nonlinear Sch\"odinger equation,
Modulation instability, Asymmetric turbulence, Rogue waves

\vspace{0.1in} \noindent Mathematics Subject Classification: 35Q51, 37K10, 65M99

\vspace{-0.05in}
\baselineskip=15pt

\section{Introduction}
\label{IN}

 Weak wave turbulence (WWT) refers to a regime in which the interactions between waves are weakly nonlinear, leading to the gradual emergence of complex statistical behaviors over time~\cite{WTbook1,WTbook2,WTbook3,WTreview1,ZP22,WTreview2}.  This phenomenon typically arises in systems governed by non-integrable dispersive wave equations,  where the wave resonance manifest in subtle energy exchanges~\cite{WT1,WT2,WT3,WT4,WT5,WT6}.  One prototypical example is one-dimensional  MMT model written as~\cite{MMT}:
 \bee
i \psi_t=\left|\partial_x\right|^\alpha \psi+\left|\partial_x\right|^{\beta / 4}\left(\left |\left| \partial_x\right|^{\beta / 4} \psi\right|^2\left|\partial_x\right|^{\beta / 4} \psi\right),
\ene
where the envelope field \(\psi=\psi(x, t) \) is a complex-valued function, the parameter \( \beta \) determines the degree of nonlinearity in the system, \( \alpha \) governs the dispersion relation, which is given by \( \omega(k) = |k|^\alpha \) with \( \omega \) being the frequency and \( k \)  the wave number. Unlike fluid turbulence, where the power law of the energy spectrum with respect to wave number is obtained through dimensional analysis, in WWT, Zakharov invented a method called the ``Zakharov transform (ZT)"~\cite{WTbook1,WTbook2}, which can derive specific power laws and obtain two dual cascades. The standard procedure typically involves employing the Fourier transform along with the assumption of independent Gaussian amplitudes and phases, which leads to the derivation of the wave-kinetic equation (WKE)~\cite{BuWT}. Subsequently, the ZT was utilized to solve the equation analytically. A critical element in this approach is the existence of resonance manifolds. For the specific steps, one can refer to these books ~\cite{WTbook1,WTbook2,WTbook3}, which provides a detailed framework for deriving the WKE and applying the transform, along with the analysis of resonance manifolds.

However, the scenario significantly differs when considering integrable nonlinear dispersive equations, which represent a particularly important and distinctive type of nonlinear physical models, and possess significant theoretical value and practical relevance in the field of applied mathematics and mathematical physics, as well as related nonlinear science. The first key aspect is that the resonance manifolds will be empty~\cite{WTbook2,Zak09,WT8}, and the second one is that integrable nonlinear wave equations permit the existence of solitons, rogue waves (RWs), and other strongly nonlinear wave structures, which are characteristic of exact solutions in such systems. The concept of integrable turbulence (IT) was first introduced by Zakharov~\cite{Zak09} to elucidate the failure of  WWT theory in integrable nonlinear wave systems and to reveal a {\it novel turbulence mechanism}. This the IT concept describes the emergence of randomness and its statistical properties resulting from the long-term evolution of waves in integrable systems. The proposal of IT not only transcends the limitations of traditional WWT, but also provides a new perspective for understanding the complex dynamical behaviors of nonlinear wave systems. Consequently, it has sparked extensive research interest in both theoretical and experimental fields~\cite{Ying2011,Sur14,Zak15, Zak16, Fr10, El2016,Akh16, Akh162, Gel18, El19, Gel19, Sur21, Xu24, Sun23, Sur24, Sur242,Cos14, Sur15, Sur16, Sur18, Sur182, Cos19, Sur19, Re20, Sur20,Zhong24}, mostly focusing on one-dimensional integrable nonlinear Schr\"odinger equation (NLSE), expressed as~\cite{PS83, WTbook2, Yang10, Fi15}:
\bee\label{nls}
\left\{\begin{aligned}
&i\psi_t + \psi_{xx} + |\psi|^2 \psi = 0, \\
&\psi(x,0) = \psi_0(x),
\end{aligned}\right.
\ene
serving as a fundamental model, holding significant importance in various disciplines such as quantum mechanics, nonlinear optics, fluid dynamics, deep ocean, plasma physics, and Bose-Einstein condensates (BECs).

Serving as another fundamental and widely studied nonlinear physical model with diverse applications in various fields,
the derivative nonlinear Schr\"odinger  equation (DNLSE)~\cite{DNLS1,DNLS2,DNLS3,IW77,DNLS4,WT22,Th,Th1},
\begin{equation}\label{DNLS}
i\psi_t+\psi_{xx}+ig(|\psi|^2\psi)_x=0,\quad g\in \mathbb{R}
\end{equation}
describes the evolution of wave phenomena in nonlinear and dispersive media.
  For example, it characterizes the propagation of circularly polarized nonlinear Alfv\'{e}n waves in plasmas~\cite{DNLS1,DNLS2,DNLS3,IW77,DNLS4} and describes the dynamics of weakly nonlinear electromagnetic waves in ferromagnetic~\cite{DNLS5}, anti ferromagnetic~\cite{DNLS6}, or dielectric~\cite{DNLS7} systems subjected to an external magnetic field.
 The DNLSE (\ref{DNLS}) is completely integrable and associated with the following modified Zakharov-Shabat eigenvalue problem~\cite{WT22}:
\begin{equation}\label{Lax1}
\Phi_x = X \Phi, \quad X = X(x, t; \lambda) = i \lambda^2 \sigma_3 + \lambda Q,
\end{equation}
\begin{equation}\label{Lax2}
\Phi_t = T \Phi, \quad T = T(x, t; \lambda) = -\left( 2\lambda^2 + Q^2 \right) X - i\lambda Q_x \sigma_3,
\end{equation}
where \( \Phi = \Phi(x, t; \lambda) \) is a \( 2 \times 2 \) matrix-valued eigenfunction, \( \lambda \in \mathbb{C} \) is a spectral parameter, the potential matrix \( Q = Q(x, t) \) is written as
\begin{equation}
Q(x, t) = \begin{bmatrix}
0 & \psi(x, t) \\
g \psi^*(x, t) & 0
\end{bmatrix},
\end{equation}
the asterisk (\(*\)) denotes the complex conjugate, and \( \sigma_3 \) is one of  Pauli's spin matrices given by
\begin{equation}
\sigma_1 = \begin{bmatrix} 0 & 1 \\ 1 & 0 \end{bmatrix}, \quad \sigma_2 = \begin{bmatrix} 0 & -i \\ i & 0 \end{bmatrix}, \quad \sigma_3 = \begin{bmatrix} 1 & 0 \\ 0 & -1 \end{bmatrix}.
\end{equation}
The DNLSE (\ref{DNLS}) is just the compatibility condition (or the zero-curvature condition), $
X_t - T_x + [X, T] = 0$
of system (\ref{Lax1})-(\ref{Lax2}).

  Abundant localized wave structures, such as solitons, rogue waves, and breathers, have been discovered for the DNLSE (\ref{DNLS})~\cite{Chen04,Las07,He11,Zhou12,Zhang20}. The parameter \(g\) represents the relative strength of the derivative nonlinear term. For simplicity, we can assume \(g = 1\), since the case where \(g = -1\) can be transformed into \(g = 1\) by applying the coordinate transformation \(x \to -x\).
The DNLSE (\ref{DNLS}) with \( g = 1 \) can be expressed in Hamiltonian form as follows ~\cite{WT22}:
\bee
\frac{\partial\psi}{\partial t}=\frac{\partial}{\partial x} \frac{\delta H}{\delta\psi^*},
\ene
with
\begin{equation}\label{Ha}
H\left(\psi\right)= H_k+H_4,
\end{equation}
where the kinetic energy (similar to the NLSE~\cite{Zak15,Zak16}) is
\begin{equation}\label{KE}
H_k=-\frac{i}{2} \int_{-\infty}^{\infty}\left(\psi \psi_x^{*}-\psi_x \psi^{*}\right)dx
\end{equation}
and the potential energy is
\begin{equation}\label{PE}
H_4= -\frac{1}{2}\int_{-\infty}^{\infty} |\psi|^4 dx.
\end{equation}
One can easily verify that the Hamiltonian in Eq.~(\ref{Ha}) is a conserved quantity. Another important conserved quantity is the mass given by:
\bee\label{LN}
N(\psi)=\int_{-\infty}^{\infty}|\psi|^2dx.
\ene
A key mechanism in the NLSE~(\ref{nls}) for generating IT~\cite{Zak15} refers to  modulation instability (MI, alias Benjamin-Feir instability)~\cite{Ben67, Zak13, Sur202}.  In the noise-induced MI model of NLSE~(\ref{nls}), the plane-wave solution is typically augmented by a small stochastic perturbation:
\bee\label{initial}
\psi_0(x) = 1 + \xi(x),
\ene
where \(\xi(x)\) represents a small noise with \(\langle|\xi|^2\rangle \ll 1\) and zero average, \(\langle\xi\rangle=0\). Here, \(\langle \cdot \rangle\) denotes the averaging over multiple realizations of the noise. Mathematically, the average is given by
\bee
\langle \xi \rangle = \int \xi p(\xi) d\xi = 0,
\ene
where \(p(\xi)\) is the probability density function (PDF) of noise \(\xi\).   It should be noted that for the NLSE, the initial condition with $\psi_0(x)$ is in the MI region. Under the influence of MI, the initial plane wave evolves into a chaotic wave field, which can be characterized by robust statistical metrics~\cite{Fr10, Zak15, Akh16, Akh162, Sur19, Gel19, Sun23, Sur24, Sur242}. In such a steady-state wave field, various sophisticated localized wave structures emerge~\cite{Zak13, Na16}, increasing the complexity and richness of wave dynamics. The interactions and collisions among these structures lead to the appearance of high-amplitude waves, commonly referred to as rogue waves (RWs)~\cite{Ro07, Akh092, Akh093}.

The study of RWs in chaotic wave fields has become a prominent area of research in nonlinear science. RWs were first observed in deep ocean, where they manifest as unusually large amplitude waves that appear unexpectedly~\cite{PS83,Fe08, Dy08, Ch11}. Characterized by tall peaks, deep troughs, and transient behavior, RWs typically last only a few seconds~\cite{Kh03, Ak10, On13,Xiao2013, Toffoli2015, Li2021, Tang2022, Mendes2023, Zhang2024,Mendes2025}. Their study has extended beyond oceanography to various other domains, demonstrating their universal presence. For example, rogue wave-like phenomena have been observed in nonlinear optical fibers~\cite{Ro07, Ki10}, quantum fluids in superfluid mechanics~\cite{superfluid}, and in plasma physics~\cite{Langmur, RW_plasma}, highlighting the widespread relevance of RWs. The RW phenomenon also appears in other fields, such as BECs~\cite{rw4, MI53}, Alfv\'{e}n waves~\cite{rw5}, and even finance~\cite{MI52}. RW generation can be driven by a variety of mechanisms, ranging from simple linear effects to complex nonlinear interactions within non-zero backgrounds. One approach to understanding RW generation is through the study of MI and IT \cite{Zak15, Akh16, Akh162, Kh03, Ak10, On13, Lak15, Arm15, Dud14, Chen2018, Rao2018, Sun21, yan21, Li2024}.

Inspired by the NLSE, a natural issue arises: does MI in the DNLSE~(\ref{DNLS}) lead to the formation of turbulence? In this case, how does the resulting turbulence differ from the one observed in the NLSE~(\ref{nls})~\cite{Zak15,Zak16}?
In this paper, we aim to systematically investigate  turbulence phenomena within the framework of the DNLSE (\ref{DNLS}), along with the formation of RWs. To achieve this aim, we perform an extensive set of numerical simulations of the DNLSE (\ref{DNLS}), using the initial condition provided in Eq.~(\ref{IC}). The key contributions of this paper can be summarized as follows:
\begin{itemize}
 \vspace{0.05in}   \item {} We carefully explore the relationship between MI and RWs in the DNLSE (\ref{DNLS}), and compared the RWs generated by the MI of plane waves with the exact one.

   \vspace{0.05in}   \item {} We broaden the concept of IT to encompass the DNLSE (\ref{DNLS}), offering a thorough examination of IT within this framework. Through various statistical measures, we  identify the occurrence of stationary IT in the DNLSE (\ref{DNLS}).

   \vspace{0.05in}   \item {} We find that, unlike the symmetric turbulence observed in the NLSE, the turbulence in the DNLSE exhibits anisotropy, with the power-law behavior varying in different regions.
\end{itemize}

 \vspace{0.05in} The rest of this paper is arranged as follows. In Sec.~\ref{MIRW}, we derive the wave number range satisfied by the unstable plane waves using MI theory, and examine their relationship with the exact RWs in the DNLSE (\ref{DNLS}). In Sec.~\ref{ITEV}, we carefully analyze the evolution of various statistical indices during the onset of turbulence, as well as the asymptotic steady state. Finally, some conclusions and discussions are presented in Sec.~\ref{CON}.

\section{Modulation instability and rogue waves}
\label{MIRW}

Prior to investigating IT, it is essential to first assess the MI of the plane wave solutions~\cite{Ben67, Zak13, Sur202}, thereby facilitating the generation of IT. A direct verification shows that the DNLSE~(\ref{DNLS}) admits the following
plane-wave solutions:
\begin{equation}\label{CW}
\psi(x,t)=Ae^{i(kx-wt)}, \quad w=k^2+A^2k,
\end{equation}
where $A>0$ is the constant amplitude, and $k\in\mathbb{R}$  is the wave-number, and $w$ the frequency of the plane wave.
We commence with a rigorous examination of the linear stability of plane wave solutions Eq.~(\ref{CW}) to the DNLSE~(\ref{DNLS}), exploring their susceptibility to perturbations via adding small disturbance as $\psi(x,t)=\left(A+\epsilon(x,t)\right)e^{i(kx-wt)}$, with $|\epsilon(x,t)| \ll 1$. Upon substituting into Eq.~(\ref{DNLS}) and linearizing with respect to $\epsilon(x,t) $, we derive the governing equation for  $\epsilon(x,t)$:
\begin{equation}\label{Pe}
i\epsilon_t+\epsilon_{xx}+i(2k\epsilon_x+2A^2\epsilon_x+A^2\epsilon^*_x)-kA^2(\epsilon+\epsilon^*)=0.
\end{equation}
Let the perturbation $\epsilon(x,t)$ be assumed to have a solution in a linear superposition of plane waves:
\begin{equation}\label{Pcw}
\epsilon(x,t)=\delta_1e^{i\left(Qx-\Omega t\right)}+\delta_2^*e^{-i\left(Qx-\Omega^* t\right)},
\end{equation}
where $\delta_\ell\, (\ell=1,2),Q,\,\Omega$ represent the constant amplitudes, wave-number, as well as frequency, respectively. Furthermore, it is readily apparent that MI arises if the imaginary part of the disturbance frequency \(\Omega\) is nonzero.  By substituting Eq.~(\ref{Pcw}) into Eq.~(\ref{Pe}), we derive that the necessary and sufficient condition for the existence of a nontrivial solution $(\left(\delta_\ell,,\ell = 1, 2\right) \neq 0)$ is that the determinant of the coefficient matrix $M$ vanishes, where
\bee
M = \begin{pmatrix}
\Omega+\omega-(k+Q)^2-2A^2(Q+k) & -A^2(k+Q) \\[1em]
-A^2(k-Q) & -\Omega+\omega-(k-Q)^2-2A^2(k-Q)  \\
\end{pmatrix}.
\ene
Therefore, the necessary and sufficient condition for the instability of the plane wave given by Eq.~(\ref{CW}) is:
\begin{equation}\label{der}
\Delta = A^4 + 2A^2k + Q^2 < 0.
\end{equation}
Moreover, for the plane wave with \( k < -\frac{A^2}{2} \), MI manifests, and the corresponding unstable growth rate (i.e., the imaginary part of \(\Omega\)) is given by:
\begin{equation}\label{Condi}
G=Q\sqrt{-A^4-2A^2k-Q^2}.
\end{equation}
It is straightforward to determine that the maximum instability (with maximum absolute value of $G$) occurs when the disturbance wave number is given by:
\begin{equation}
Q_{\text{max}} = \pm \sqrt{\frac{-A^2(2k + A^2)}{2}},
\end{equation}
 and the maximum growth rate
of the instability,
\begin{equation}\label{Mai}
G_{0} = \frac{-A^2(2k + A^2)}{2}.
\end{equation}
Therefore, the characteristic length of the instability is \( \lambda = 2\pi/ |Q_{\text{max}}| \), and the characteristic time is \( \tau = 1/G_0 \).
\begin{figure}
\centering
{\scalebox{0.5}[0.5]{\includegraphics{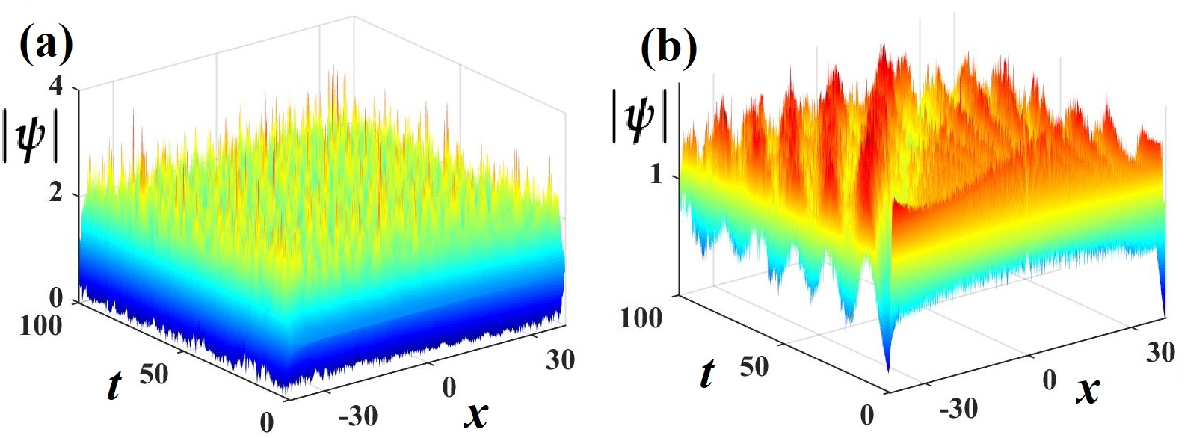}}}
\caption{ A typical evolution of the turbulent wave field \( |\psi(x,t)| \) for a single realization corresponds to the  parameters at MI region  (\textit{a})    with  \( k = - 3 \), and at MS region (\textit{b}) with \( k = -1/4 \).}
\label{MI}
\end{figure}

\begin{figure}
\centering
{\scalebox{0.5}[0.5]{\includegraphics{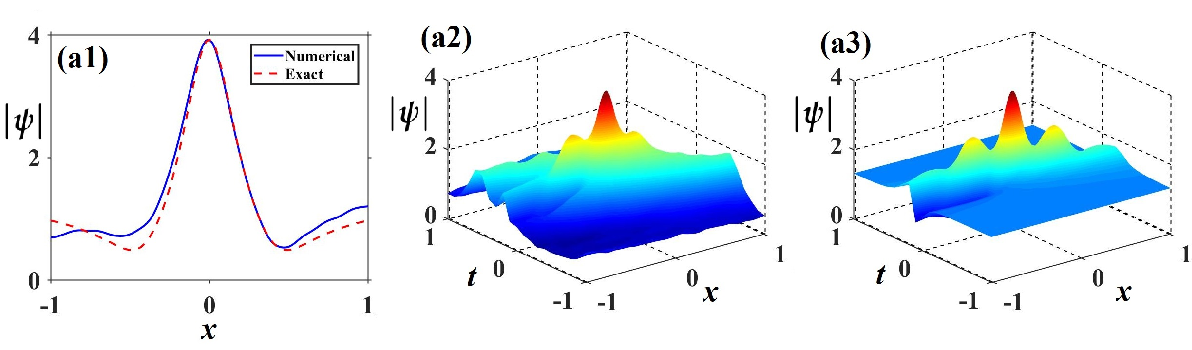}}}
\caption{ The largest RWs  detected in the realization corresponding to figure~\ref{MI}(a). (\textit{a1}) The comparison between  the largest RW  detected in the realization and the cross-profile of exact RW solution, where the blue solid and red dashed lines represent the numerical and exact ones, respectively.  The space-time representation near the location where the maximum amplitude is attained of numerical realization at (\textit{a2}) and exact solution at (\textit{a3}). To enhance visualization, we translate the point of maximum amplitude to the coordinates \((x,t) = (0,0)\).  }
\label{RW}
\end{figure}
Now, we proceed with a numerical verification of MI and its relationship with RWs. Initially, we assign the amplitude of the plane wave parameter as \( A = 1 \). Hence, MI occurs at
\bee
k<-\frac{1}{2}.
\ene
The numerical results are shown in figure~\ref{MI}, with the parameter \( k \) being located in the MI region at \( k =- 3 \) in panel (\textit{a}), and in the modulation stable (MS) region at \( k = -1/4 \) in panel (\textit{b}).
The numerical results provide a strong validation for the theory. We observe that in the MI region (see figure~\ref{MI}(\textit{a})), the amplitude of the plane wave increases significantly due to the instability, reaching values greater than 3. In contrast, in the MS region (see figure~\ref{MI}(\textit{b})), the amplitude does not experience significant growth over time. It should be noted here that the linear instability we consider does not cause the ill-posedness of the solution \cite{BIO}.

Subsequently, we investigate the relationship between MI and RWs.  In figure~\ref{RW}, we present  the largest extreme RW event detected in the realization corresponding to figure~\ref{MI}(\textit{a}).  The comparison between the detected one and exact RW with parameters $\alpha=0.95,\beta=0.5$ (see Eq.~(55) in \cite{He11}) is displayed in figure~\ref{RW}(\textit{a1}),  which exhibits a noticeable level of consistency.
The space-time representation near the location of the maximum amplitude, shown for the numerical realization in figure~\ref{RW}(\textit{a2}) and the exact solution in figure~\ref{RW}(\textit{a3}).  Due to the instability induced by MI, the localization of the RW in figure~\ref{MI}(\textit{a2}) is slightly less precise; however, a general agreement with the exact RW can still be observed.

\section{Formation of asymmetric integrable turbulence}
\label{ITEV}
\begin{figure}
\centering
{\scalebox{0.5}[0.5]{\includegraphics{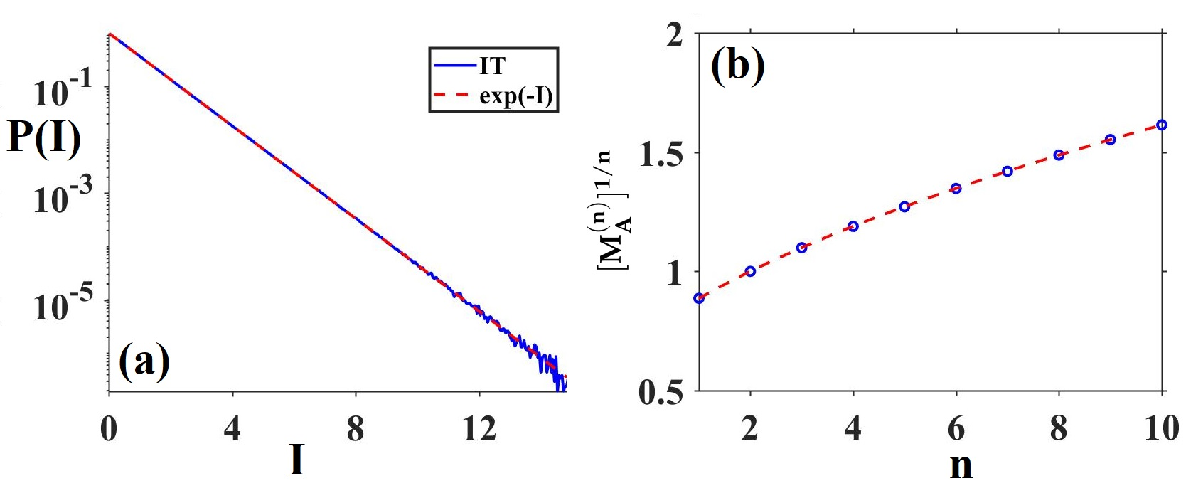}}}
\caption{(\textit{a}) The  PDF of intensity $I=|\psi|^2$, where the blue solid line represents asymptotic state of IT, while the red dashed line denotes the exponential distribution $\exp(-I)$. (\textit{b}) The asymptotic values of the moments \( M^{(n)}_A \), where \( n = 1, \dots, 10 \), are represented by blue circles. These values are compared with the Rayleigh prediction, \( \Gamma\left(\frac{n}{2} + 1\right) \), which is shown as a dashed red line.   }
\label{MP}
\end{figure}

First, building upon the stability analysis presented in the preceding section, we examine the DNLSE~(\ref{DNLS}), with the initial conditions specified as:
\begin{equation}\label{IC}
\psi(x,0)=e^{-3ix}\left(1+\text{noise}\right),
\end{equation}
that is, $A=1,\, k=-3$ in the plane-wave solution in Eq.~(\ref{CW}), which lies within the parameter regime of MI. Note that the initial conditions here can also be adjusted, as long as the parameters remain within the range of MI. The initial condition contains a random perturbation in the form of small-amplitude Gaussian noise.
Considering IT, we will analyze it using the following different statistical indexes in this section. And the IT  will be characterized from two distinct perspectives: first, by examining the steady-state values of its statistical metrics, and second, by analyzing the dynamical evolution of these metrics as they approach to their equilibrium states. For details of the specific numerical algorithm, readers are referred to Appendix~\ref{NM}.
\subsection{Moments and probability density function}

\begin{figure}
\centering
{\scalebox{0.5}[0.5]{\includegraphics{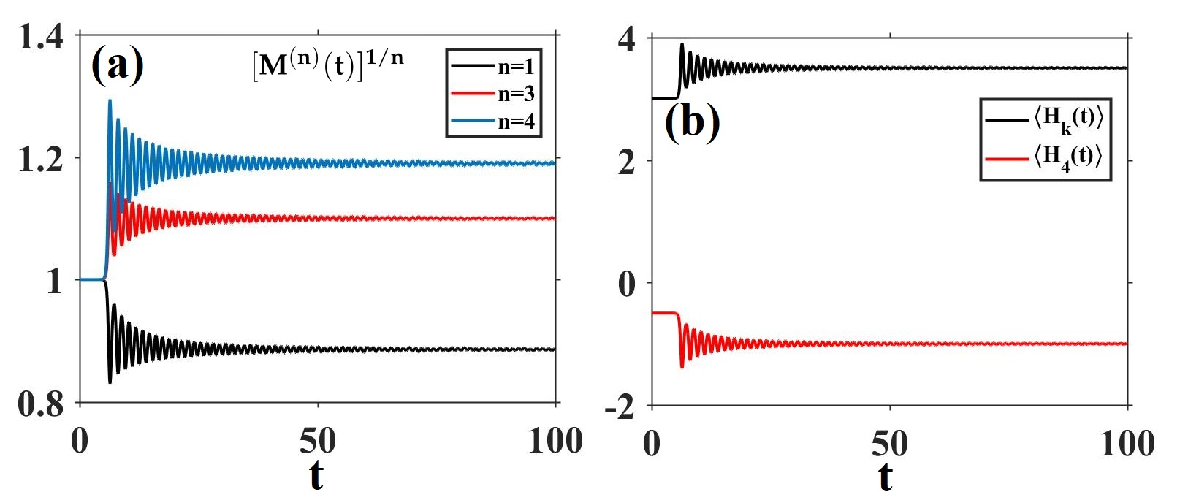}}}
\caption{(\textit{a}) The evolution of $n$-th moment $\left[M^{(n)}(t)\right]^{1/n}$  (see Eq.~(\ref{Moment})) ($n=1,2,3$) with time $t$. (\textit{b}) The evolution of ensemble-averaged kinetic energy $\langle H_k(t) \rangle$ Eq.~(\ref{KE}) and potentia  energy $\langle H_4(t)\rangle$  Eq.~(\ref{PE}) with time $t$.   }
\label{EMP}
\end{figure}

An essential statistical measure in this context is the fourth-order moment, which is directly linked to kurtosis, derived from the initial conditions given in Eq.~(\ref{IC}). This quantity provides crucial insights into the distribution's peakedness and tail behaviors, which are vital for understanding the formation of rogue waves in the turbulent regime of nonlinear wave dynamics~\cite{Zak15,Zak16}. The general expression for the \(n\)-th moment is given by:
\begin{equation}\label{Moment}
M^{(n)}(t) = \left\langle \frac{1}{L} \int_{-L / 2}^{L / 2} |\psi(x, t)|^n \, dx \right\rangle.
\end{equation}
It is worth noting that the second moment, \(M^{(2)}(t) \equiv M^{(2)}(0) \approx 1\), reflects  the DNLSE~(\ref{DNLS}), which holds true throughout the evolution. Indeed, Eq.~(\ref{Moment}) can be rewritten in the form of  PDF as shown below:
\begin{equation}\label{Co}
M^{(n)}(t)=\int_0^{+\infty}|\psi|^n \mathcal{P}(|\psi|, t) d|\psi|.
\end{equation}
If the PDF of wave amplitudes coincides with Rayleigh distribution:
\begin{equation}\label{RAD}
P_{Ra}\left(|\psi|\right)=2|\psi|e^{-|\psi|^2},
\end{equation}
then it can be shown that the \( n \)-th moment, \( M_{Ra}^{(n)}(t) \), is given by \( \Gamma\left(\frac{n}{2} + 1\right) \). By performing a variable substitution \( I = |\psi|^2 \), the PDF of the intensity \( I \) can be obtained as $P_{Ra}\left(I\right)=\exp(-I)$, which is called  the exponential distribution.

\begin{figure}
\centering
{\scalebox{0.45}[0.45]{\includegraphics{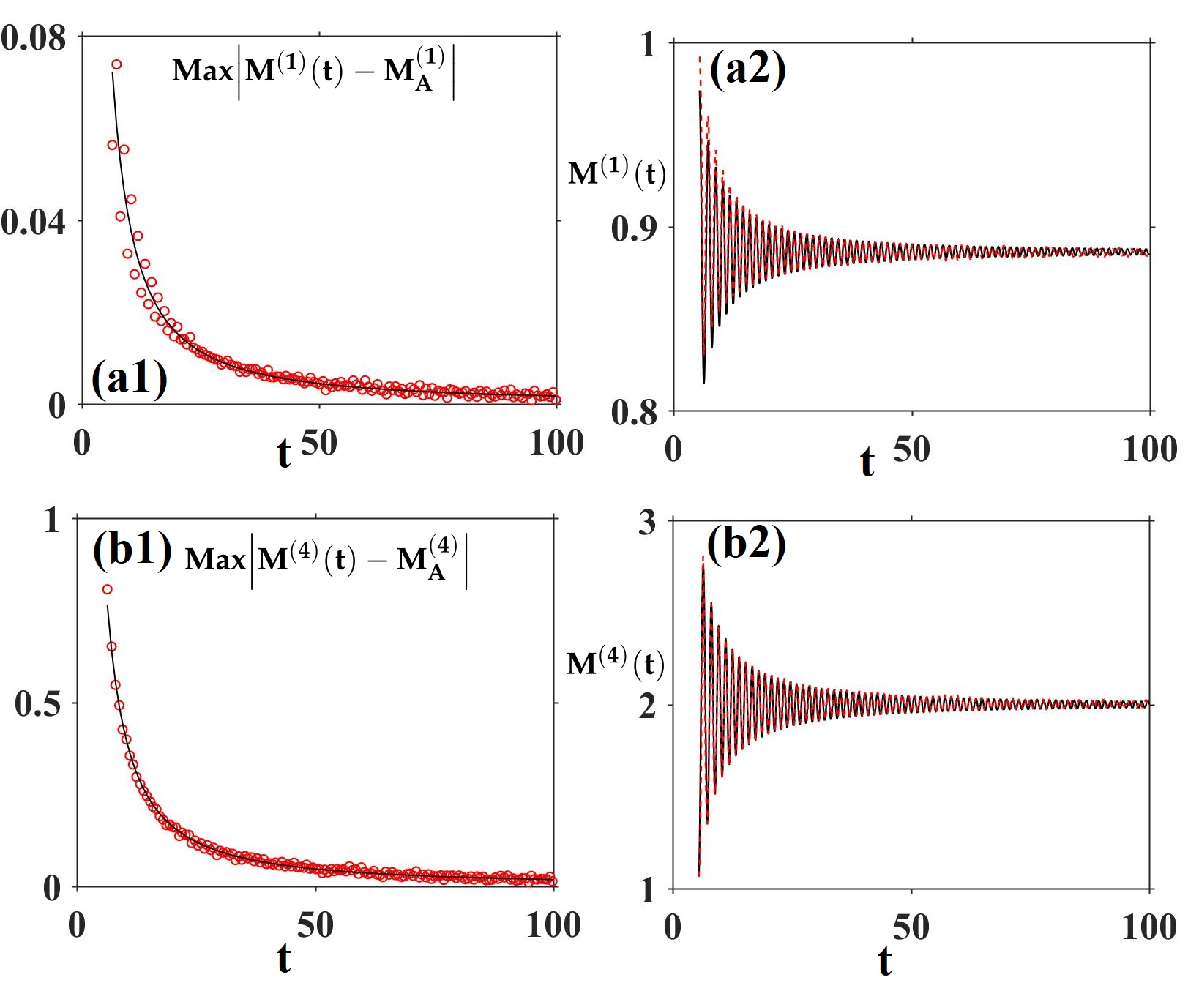}}}
\caption{(\textit{a1}) The magnitude of the deviations of the extreme of \( M^{(1)}(t) \) from its asymptotic value \( M_A^{(1)} \), as a function of time \( t \). (\textit{a2}) The evolution of the moment \( M^{(1)}(t) \) (solid black line) is fitted by the function \(  M_A^{(1)} + \frac{a_1}{t^{\alpha_1}} \sin \left(b_1 t+c_1/t^{\beta_1}+\theta_0^{(1)}\right)\), with the parameters $M_{A}^{1}\approx0.886$, $\{a_1,\alpha_1,b_1,c_1,\beta_1,\theta_0^{(1)}\}\approx$ $\{0.89,1.36,4.89,51.41,0.78,-44.59\}$ (dashed red line). (\textit{b1}) The same as (a1), expect \( M^{(1)}(t) \) is changed into \( M^{(4)}(t) \). (\textit{b2}) The same as (a2), expect \( M^{(1)}(t) \) is changed into \( M^{(4)}(t) \), as well as  the parameters $M_{A}^{4}\approx2, \{a_2,\alpha_2,b_2,c_2,\beta_2,\theta_0^{(2)}\}\approx\{9.39,1.35,4.89,51.06,0.80,-41.46\}.$     }
\label{FM}
\end{figure}

As depicted in figure~\ref{MP}(\textit{a1}), the asymptotic PDF of the intensity, \( P(I) \), matches the exponential PDF, $\exp(-I)$. This observation leads to a rather unexpected conclusion: although the DNLSE~(\ref{DNLS}) is inherently nonlinear, its asymptotic PDF is indistinguishable from that of a wave field governed by a linear equation. To further corroborate this finding, we computed the asymptotic values of the moments $M^{(n)}_{A}$, and discovered that they are in good agreement with the Rayleigh predictions  for \( n = 1, ..., 10 \), as illustrated in figure~\ref{MP}(\textit{b}). Furthermore, we discovered the following relationship between the ensemble
average potential energy \( H_4 \) and the fourth-order moment as
\bee\label{Rela}
\langle H_4\rangle=-\frac{1}{2}M^{(4)}_{A}=-1.
\ene
Therefore, the asymptotic value of ensemble
average kinetic energy $\langle H_k\rangle=\frac{7}{2}$ since the conservation of ensemble
average Hamiltonian  $\langle H\rangle=\langle H_k\rangle+\langle H_4\rangle=\frac{5}{2}.$ The nonlinearity  degree  in asymptotic state of IT can be estimated using the following parameter:
\bee\label{Dn}
R=\frac{|\langle H_4\rangle|}{\langle H_k\rangle}=\frac{2}{7}.
\ene
In the case of WWT, where \( |R| \ll 1 \), the Rayleigh PDF naturally arises as the expected outcome. However, for the DNLSE, we observe a regime of "slightly strong" turbulence in the asymptotic state, with \( R = 2/7 \). This suggests a shift from the WWT regime, indicating a significant increase in nonlinearity as the system evolves.

The relation given in Eq.~(\ref{Rela}) is particularly noteworthy. Referring to Eq.~(\ref{PE}), we can write:
$$
-\langle H_4 \rangle = \frac{1}{2} \sum_{k_1, k_2, k_3, k_4} \langle \psi_{k_1} \psi_{k_2} \psi_{k_3}^* \psi_{k_4}^* \rangle \delta_{k_1+k_2-k_3-k_4},
$$
where $\psi_{k_i}$ is the Fourier transform as defined in Eq.~(\ref{SD}), \(\delta_k\) is the Kronecker delta function defined as:
\bee
\delta_k = \begin{cases}
1, & \text{if } k = 0, \\
0, & \text{if } k \neq 0.
\end{cases}
\ene
The four-point correlation, or four-wave momentum, can be rewritten as ~\cite{WT2,Zak15,Zak16}:
\bee
\langle \psi_{k_1} \psi_{k_2} \psi_{k_3}^* \psi_{k_4}^* \rangle = I_{k_1} I_{k_2} \left( \delta_{k_1-k_3} \delta_{k_2-k_4} + \delta_{k_1-k_4} \delta_{k_2-k_3} \right) + J_{k_1, k_2, k_3, k_4},
\ene
where \(J_{k_1, k_2, k_3, k_4}\) represents the cumulant, $I_{k_i}$ denotes the wave-action spectrum (see Eq.~(\ref{Wa})). Noting that the sum of \(I_k\) terms is approximately the mean mass, \(\sum I_k = \langle N \rangle \approx 1\), we obtain the following approximation:
\bee
-\langle H_4 \rangle \approx 1 + \frac{1}{2} \sum_{k_1, k_2, k_3, k_4} J_{k_1, k_2, k_3, k_4} \delta_{k_1+k_2-k_3-k_4}.
\ene
When we combine this with the relation from Eq.~(\ref{Rela}), we get:
\bee
\left| \sum_{k_1, k_2, k_3, k_4} J_{k_1, k_2, k_3, k_4} \delta_{k_1+k_2-k_3-k_4} \right| \ll 1.
\ene
This inequality suggests that, in the asymptotic limit of the turbulent state, the cumulant \(J_{k_1, k_2, k_3, k_4}\) tends to zero. Therefore, the fact that \(\langle H_4 \rangle \approx -1\) strongly implies that the turbulence is Gaussian in nature in its stationary state.

To investigate the evolution towards the asymptotic turbulent state, as displayed in figure~\ref{EMP}, we examine the moments \( M^{(n)}(t) \),  the ensemble-averaged kinetic energy \( \langle H_k \rangle \), potential energy \( \langle H_4 \rangle \). Initially, up to \( t \sim 5\), the perturbations to the condensate are minimal, resulting in only slight variations from their initial values: \( M^{(n)} \approx 1 \), \( \langle H_k \rangle \approx 3 \), and \( \langle H_4 \rangle \approx -1/2 \). At around \( t \sim 5 \), the MI transforms into its nonlinear phase, marking a significant shift. During this phase, the moments start to oscillate about their asymptotic Rayleigh values $\Gamma(n/2+1)$, with the kinetic energy stabilizing at approximately 7/2, and the potential energy around $-1$.

Remarkably, the moment \( M^{(1)}(t) \) exhibits in-phase oscillations with the potential energy \( \langle H_4 \rangle \), while it oscillates out of phase with the higher-order moments \( M^{(n)}(t) \) (for \( n \geq 3 \)) and the kinetic energy \( \langle H_k \rangle \) during the transition from the nonlinear stage of MI to the asymptotic state. Consequently, the temporal positions of the local maxima and minima of \( M^{(1)}(t) \) and \( \langle H_4 \rangle \) coincide, aligning precisely with the local minima and maxima of \( M^{(n)}(t) \) (for \( n \geq 3 \)) and \( \langle H_k\rangle \), respectively. This intricate phase relationship reveals the underlying dynamics governing the evolution of turbulence in this system. In fact, the result can be derived from a relatively straightforward perspective. First, it is important to note that \( \Gamma(n/2 + 1) \leq 1 \) holds if and only if \( n \leq 2 \). Since the initial values of all \( n \)-th moments are unity, it follows that \( M^{(1)}(t) \) initially decreases, while the higher-order moments \( M^{(n)}(t) \) (for \( n \geq 3 \)) increase. A similar analysis can be applied to the kinetic energy \( \langle H_k \rangle \) and the potential energy \( \langle H_4 \rangle \), with the crucial observation that the Hamiltonian is conserved throughout the process.

We investigate the time dependence of the oscillations using the examples of the moments \( M^{(1)}(t) \) and \( M^{(4)}(t) \). figure~\ref{FM}(\textit{a1}) reveals that the oscillatory amplitude of \( M^{(1)}(t) \), defined as the absolute deviation of its extreme from the asymptotic  value \( M_A^{(1)} \), exhibits a power-law decay governed by the scaling relation \( a_1/t^{\alpha_1} \), with a numerically determined prefactor \( a_1 = 0.89  \) and the power $\alpha_1=1.36$.  Notably, the oscillation period undergoes a progressive reduction over time, decreasing from \( T \sim 1.5 \) at \( t \sim 10 \) to \( T \sim 1.3 \) at \( t \sim 100 \). This temporal compression of the period is reminiscent of a nonlinear phase shift. Hence such behavior suggests that the temporal evolution of \( M^{(1)}(t) \) can be effectively captured by a tailored functional form as:
\begin{equation}\label{Fit}
M^{(1)}(t) \approx M_A^{(1)}+\frac{a_1}{t^{\alpha_1}} \sin \left(\theta^{(1)}\left(t\right)\right), \quad \theta^{(1)}(t)=b_1 t+\theta_{n l}^{(1)}(t)+\theta_0^{(1)},
\end{equation}
where $\theta^{(1)}(t)$ is the phase due to the existence of periodic oscillation, including frequency $b_1$, nonlinear phase shift $\theta_{n l}^{(1)}(t)$ as well as initial phase $\theta_0^{(1)}$. It is worth mentioning that the nonlinear phase shift is also a power function of time $t$, i.e., $\theta_{n l}^{(1)}(t)=c_1/t^{\beta_1}$ with $\beta_1=0.78$.  Indeed, the parameters \( a_1 \) and \( \alpha_1 \) can be determined by identifying the extreme of \( |M^{(1)}(t) - M_{\text{A}}^{(1)}| \), as clearly illustrated in figure~\ref{FM}(\textit{a1}). Then the parameters $\{b_1,c_1,\theta_0^{(1)}\}$ can be determined by the following conditions:
\bee
\theta\left(t_{\text{max}}^{m}\right)=b_1 t_{\text{max}}^m+\theta_{n l}(t_{\text{max}}^{m})+\theta_0=\frac{3\pi}{2}+2\pi m
\ene
with $m=0,1,2$, and $t_{\text{max}}^m$ represents the deviations of local maximums  of $M^{(\text{1})}(t)$ from it's asymptotic value $ M^{(1)}_{A}$. Then the parameters can be identified as $\{4.89,51.41,-44.59\}$. The evolution of $M^{(\text{1})}(t)$ and the corresponding fitting function Eq.~(\ref{Fit}) are depicted in figure~\ref{FM}(\textit{a2}), which reveals a strong alignment.
\begin{figure}
\centering
{\scalebox{0.5}[0.5]{\includegraphics{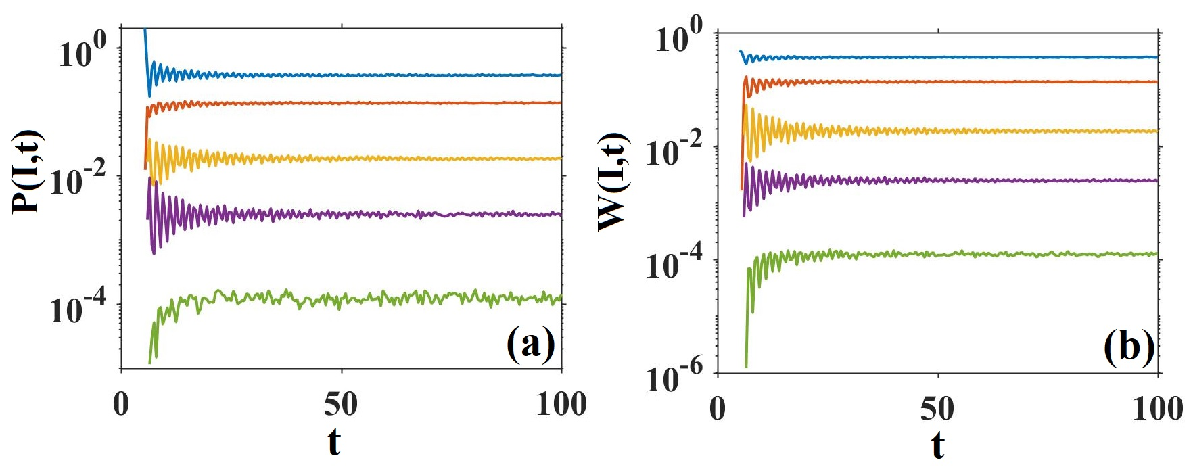}}}
\caption{(\textit{a1})  Evolution of  PDF  $P(I,t)$ for specific $I$, $I = 1$
(blue), $I = 2$ (red), $I=4$ (yellow), $I = 6$ (purple), $I = 9$ (green). (\textit{a2}) Similar with panel (a1) expect the PDF $P(I,t)$ is replaced by cumulative probability \( W(K, t) \).   }
\label{PDFSingle}
\end{figure}

Analogously, we could investigate the variations of $M^{(\text{4})}(t)$ and the associated fitting function as:
\begin{equation}\label{Fit2}
M^{(4)}(t) \approx M_A^{(4)}+\frac{a_2}{t^{\alpha_2}} \sin \left(\theta^{(2)}\left(t\right)\right), \quad \theta^{(2)}(t)=b_2 t+\theta_{n l}^{(2)}(t)+\theta_0^{(2)},
\end{equation}
with $\theta_{n l}^{(2)}(t)=c_2/t^{\beta_2}$ with $\beta_2=0.8$. It is evident that $\beta_1$ and $\beta_2$ exhibit remarkable similarity, and this discrepancy could potentially be rectified with a greater volume of ensembles.
The parameters $\{a_2,\alpha_2,b_2,c_2,\theta_0^{(2)}\}$ are numerically found as $\{9.39,1.35,4.89,51.06,-41.46\}$. It is observed that the parameters $\{\alpha_2, b_2, c_2\}$ are very close to $\{\alpha_1, b_1, c_1\}$, which is natural. However, the difference between $a_2$ and $a_1$ is more pronounced. This can be attributed to the fact that $M^{(4)}(t)$ deviates significantly from its asymptotic steady-state value when initially entering the nonlinear MI phase, in contrast to $M^{(1)}(t)$. Meanwhile, we have
\bee
\theta_0^{(1)} \approx \theta_0^{(2)} +\pi,
\ene
which arises from their anti-phase evolution, resulting in the observed effect. The results are presented in figure~\ref{FM}(\textit{b1,b2}). As for the evolution of other moments, the ensemble-averaged kinetic energy \( \langle H_k \rangle \) as well as the potential energy \( \langle H_4 \rangle \) can be similarly obtained. Note that in the case of $A=1,k=-3$, we have
\bee
Q_{\text{max}}=\pm\sqrt{5/2},\quad G_0=5/2.
\ene
Similar with the work in~\cite{Zak15}, we have $b_1=b_2\approx 2G_0$, and we find that the frequency $b_1$ or $b_2$ should coincide with 5 if we consider lager ensembles.

\begin{figure}
\centering
{\scalebox{0.5}[0.5]{\includegraphics{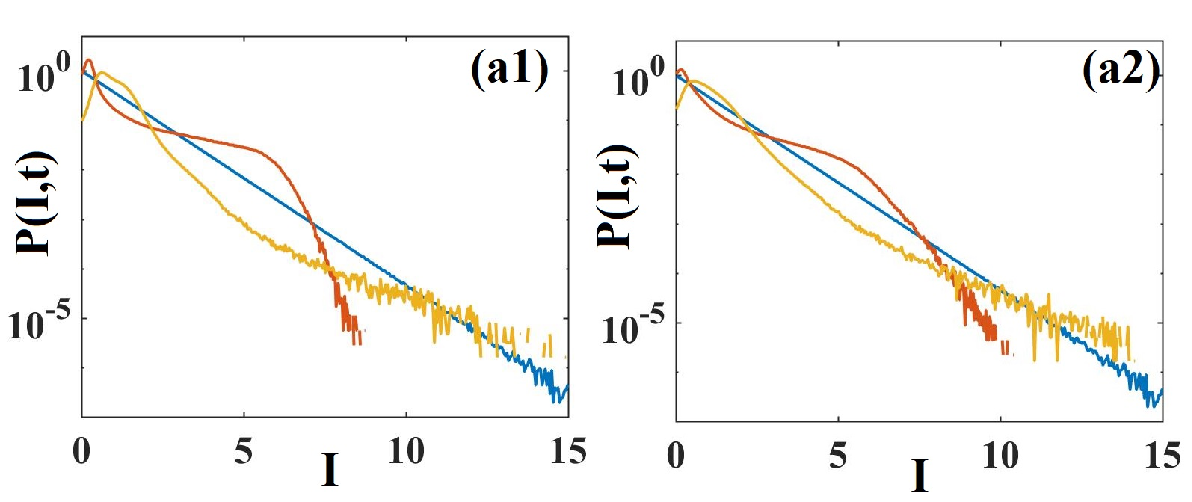}}}
\caption{(\textit{a1})  The  PDF of intensity,  $P(I,t)$, at $t=6.4$, corresponding to the first local maximum of the ensemble-averaged potential energy modulus $|\langle H_4\rangle|$ (red line); at \(t = 7.3\), corresponding to the first local minimum of $|\langle H_4\rangle|$ (yellow line); and as the asymptotic spatial correlation function (blue line).  (\textit{a2})  The  PDF of intensity,  $P(I,t)$, at $t=8.1$, corresponding to the second local maximum of the ensemble-averaged potential energy modulus $|\langle H_4\rangle|$ (red line); at \(t = 8.9\), corresponding to the second local minimum of $|\langle H_4\rangle|$ (yellow line); and as the asymptotic spatial correlation function (blue line).   }
\label{PDFDouble}
\end{figure}

Next, we will carefully investigate the time evolution of $P(I,t)$ for the specific $I = |\psi|^2$, where the results for $I=\{1,2,4,6,9\}$ are exhibited in figure~\ref{PDFSingle}(\textit{a1}). It can be observed that the evolution is in-phase or anti-phase with that of ensemble-averaged potential energy modulus $|\langle H_4\rangle|$ (see figure~\ref{FM}(\textit{b2}) but with different values). We opt for the absolute value of the ensemble-averaged potential energy, as its local maximal and minimal correspond to the moments of  strongest and  weakest effects, respectively. Then in phase  means that the statistic index  take local maximal values at the local maximums of potential energy modulus \( |\langle H_4 \rangle| \), and local minimal values
at the local minimums of \( |\langle H_4 \rangle| \). The anti-phase relationship is the opposite.   It is noted that during the linear MI stage, the probability density at \( I = 1 \) is significantly greater than at other intensities. As the nonlinear MI develops, \( P(1,t) \) will approach its asymptotic value in an inversely proportional manner to \( |\langle H_4 \rangle| \). Meanwhile, \( P(2,t) \), \( P(4,t) \), and \( P(6,t) \) also approach their respective asymptotic values, in phase with \( |\langle H_4 \rangle| \). However, the evolution of \( P(9,t) \) is more erratic, as \( I = 9 \) corresponds to a rare event.

The probability of waves exceeding a certain threshold \( I > K \) is given by the cumulative probability \( W(K, t) \), which can be expressed as:
\bee\label{Cu}
W(K, t) = \int_{K}^{+\infty} P(I, t) \, dI.
\ene
Note that in case of the Rayleigh PDF in Eq.~(\ref{RAD}),  this probability takes the simple form
\bee\label{Cu2}
W(K, t) = e^{-K}.
\ene
The results for $K=\{1,2,4,6,9\}$ are summarized in figure~\ref{PDFSingle}(\textit{a2}), where similar oscillations can be observed.   For the final state, we compared our numerical results with the theory in Eq.~(\ref{Cu2}) and found a very good agreement, with an error on the order of \( 10^{-4} \).

\begin{figure}
\centering
{\scalebox{0.5}[0.5]{\includegraphics{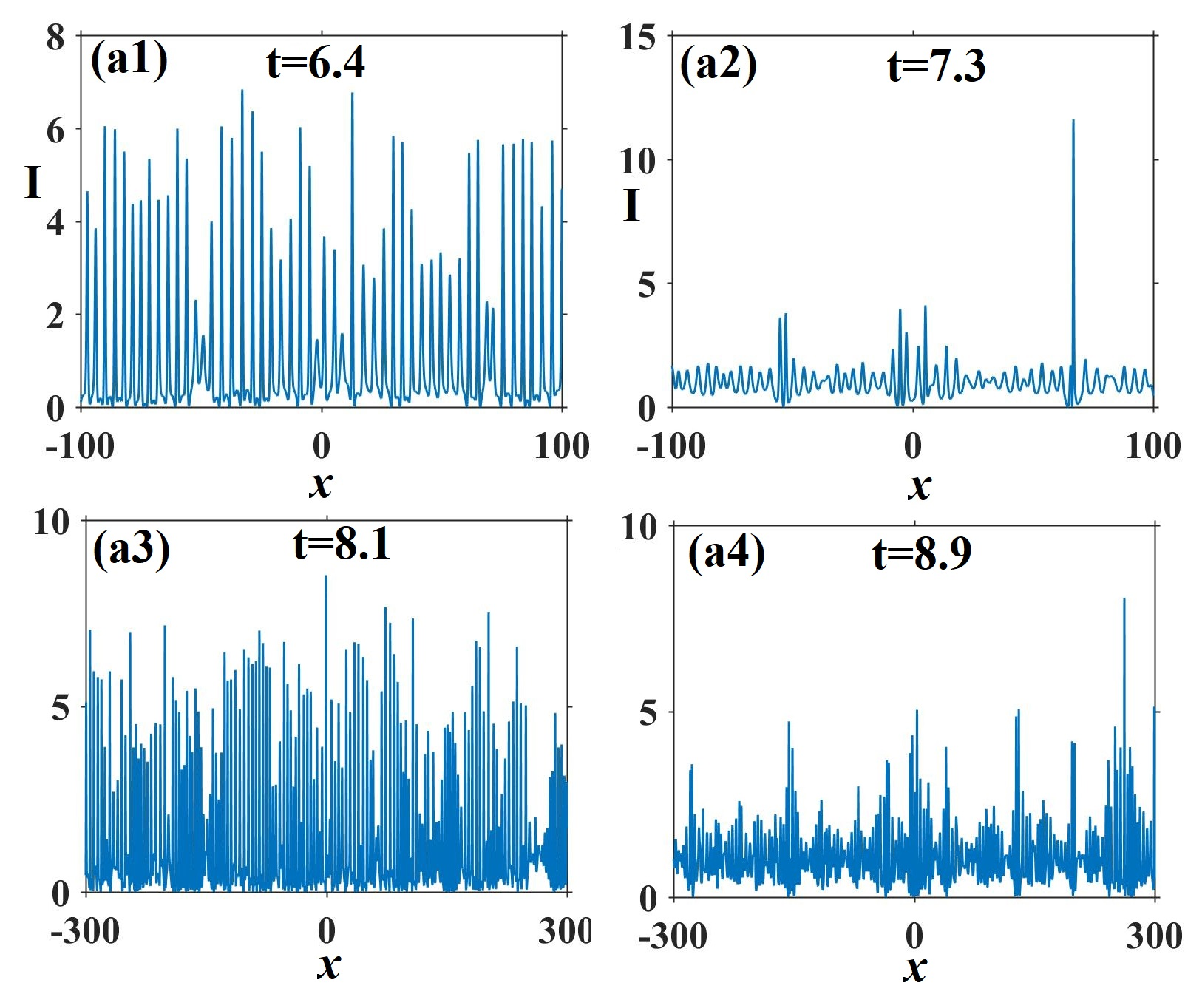}}}
\caption{Spatial distribution of the intensity $I=|\psi|^2$ at (\textit{a1}) $t=6.4$, (\textit{a2}) $t=7.3$, (\textit{a3}) $t=8.1$ as well as (\textit{a4}) $t=8.9$ for single realization.   }
\label{RWC}
\end{figure}

Next, we consider  \( P(I,t) \) in the cases of the local maxima and minima of the ensemble-averaged potential energy modulus. As depicted in figure~\ref{PDFDouble}(\textit{a1}), it is observed that at the first local maximal time (\( t = 6.4 \)), the PDF surpasses the asymptotic distribution within the interval \( I \in [3, 7] \), a region identified as the "imperfect rogue wave"~\cite{Zak15} (see also figure~\ref{PDFDouble}(\textit{a2}) at \( t = 8.1 \)). Due to the influence of MI, the "imperfect" rogue waves typically appear at the first few local maxima of \( |\langle H_4 \rangle| \). In spatial terms, these waves form a modulated lattice structure, consisting of large waves, as displayed in figure~\ref{RWC}(a1,a3) with time $t=6.4$ and $t=8.1$, which correspond to the first and second moment of local maxima. Next, we focus on the first two moments of local minimum values of \( |\langle H_4 \rangle| \), as illustrated in figure~\ref{PDFDouble}(\textit{a1,a2}). In the range of \( 12 \leq I \leq 15 \), the PDF at these instances surpasses the asymptotic PDF. These waves are uncommon occurrences, emerging in the context of a disturbed wave field (see a typical spatial distribution with $t=7.3,8.9$ at figure~\ref{RWC}(\textit{a2,a4}) ), which generally has an amplitude smaller than \( I < 4 \). Due to the influence of the PDF, during the early stages of nonlinear MI, at the moments corresponding to the local maxima of \( |\langle H_4 \rangle| \),  the wave-action spectrum \( S_k(t) \) has more peaks but smaller peak values, compared to the moments corresponding to the local maxima of \( |\langle H_4 \rangle| \). A similar phenomenon also occurs in the auto-correlation function \( g(x,t) \), which is due to the the wave field  is less correlated.

\subsection{Wave-action spectrum and auto-correlation function}

In this subsection, we aim to examine the asymptotic behavior of the wave-action spectrum \( S_k(t) \) and the auto-correlation function \( g(x,t) \), as well as their progression toward the final state. The wave-action spectrum \( S_k(t) \) can be defined as
\bee\label{Wa}
S_k(t)=\langle|\psi_k(t)|^2 \rangle.
\ene
 The quantity \(\psi_k(t)\), defined as the semi-discrete Fourier transform, differs from its continuous counterpart primarily because periodic boundary conditions are imposed by the numerical methods we use. These transforms are expressed as:
\bee\label{SD}
\begin{aligned}
\psi_k(t) &= \mathcal{F}_D[\psi(x, t)] = \frac{1}{L} \int_{-L/2}^{L/2} \psi(x, t) e^{-i k x} \, dx, \\[0.8em]
\psi(x, t) &= \mathcal{F}_D^{-1}[\psi_k(t)] = \sum_k \psi_k(t) e^{i k x}.
\end{aligned}
\ene
To distinguish them from the continuous Fourier transforms, we denote them as \(\mathcal{F}_D\) and \(\mathcal{F}_D^{-1}\). Here, the length \(L = 2m\pi\), where \(m \gg 1\) is a large integer, and the wave number is defined as \(k = 2n\pi / L\) with \(n\) being an integer.  The wave-action spectrum is the spectral density of wave action since
\bee
\langle N\rangle=\sum_k S_k(t) .
\ene
 According to Eq.~(\ref{SD}), all wave-action spectrum of the initial condition in Eq.~(\ref{IC})  is concentrated in the   $k=-3$ harmonic,
\bee
I_k= \begin{cases}1, & k=-3, \\ 0, & k \neq -3.\end{cases}
\ene

According to the theory of MI, in the linear stage of MI, the wave number $Q$ located between
\bee\label{Ra}
-\sqrt{5}\leq Q\leq\sqrt{5}
\ene
would be unstable, and grow exponentially, expect for $Q=0$, while the wave number $Q$ outside this range do not change with time.

\begin{figure}
\centering
{\scalebox{0.6}[0.6]{\includegraphics{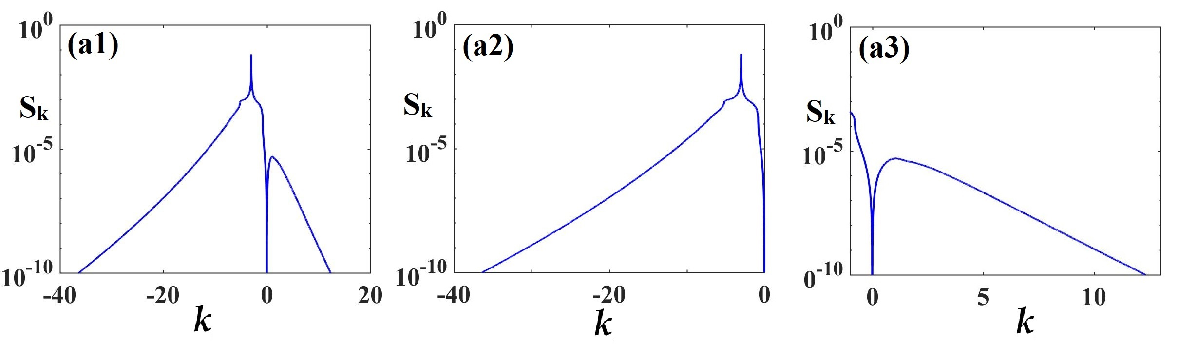}}}
\caption{The asymptotic wave-action spectrum Eq.~(\ref{Wa}) with wave number $k$ between $(-40,20)$ at (\textit{a1}). The zoomed-in images of panel (\textit{a1}) are respectively located at the vicinity of \([-40, 0]\) at (\textit{a2}), while (\textit{a3}) corresponds to \([-1, 13]\). }
\label{Sk}
\end{figure}

The asymptotic steady state of $S_k$ is shown in figure~\ref{Sk}, where certain regions exhibit power-law decay while others exhibit exponential decay. Overall, the asymptotic steady state of $S_k$ is asymmetric, a feature does not observed in the integrable turbulence of the NLS equation~\cite{Zak15,Zak16}. Before proceeding with the detailed analysis, it is important to first elucidate the underlying factors that give rise to anisotropy. This is primarily due to the selection of the wave number \(k=-3\), which is not centered within Eq.~(\ref{Ra}). Specifically, during the linear stage of MI development, the wave numbers described by Eq.~(\ref{Ra}) become excited unless \(Q=0\). In the nonlinear stage, other wavenumbers are also excited, but the majority of the wave-action spectrum remains concentrated around the linearly excited MI wavenumbers and near \(k=-3\). This results in anisotropy. In contrast, for the NLS case~\cite{Zak15,Zak16}, their chosen wave number \(k=0\) coincides with the center of the MI wave number range, leading to an symmetric result.

\begin{figure}
\centering
{\scalebox{0.5}[0.5]{\includegraphics{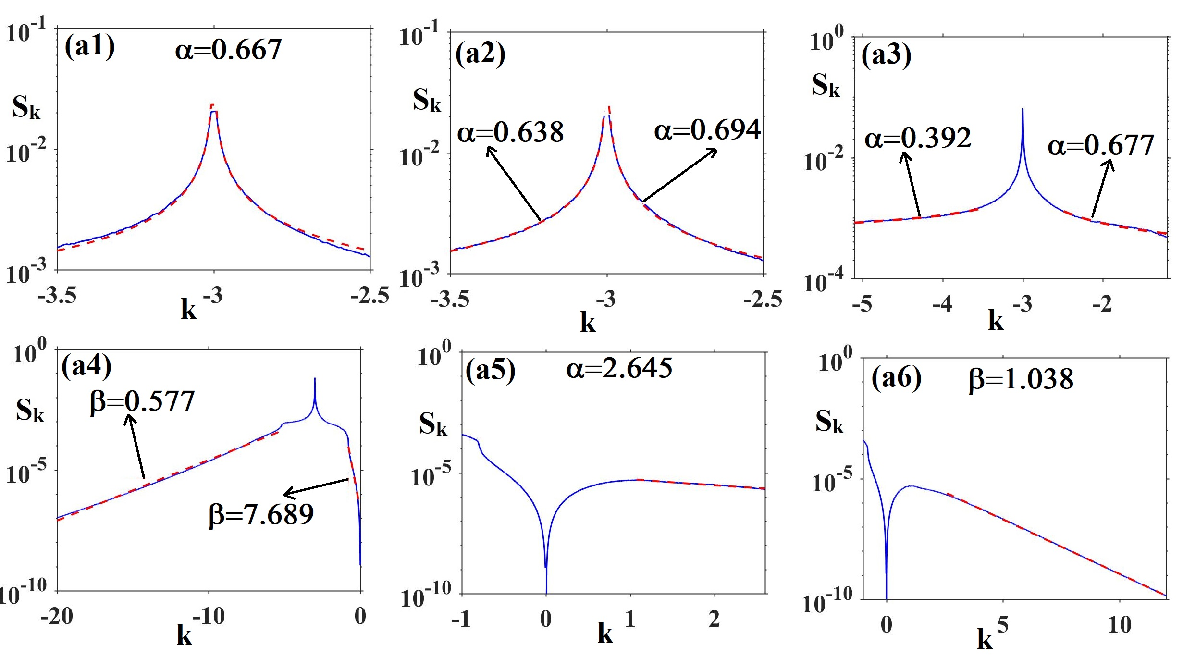}}}
\caption{The asymptotic wave-action spectrum Eq.~(\ref{Wa}) fitting by different functions in different regions. The parameters of algebraic decay Eq.~(\ref{AD}): (\textit{a1}) \(C = 9.061 \times 10^{-4}\) and \(\alpha = 0.667\); (\textit{a2}) Left part with \(C = 9.856 \times 10^{-4}\) and \(\alpha = 0.638\), right part with \(C = 8.332 \times 10^{-4}\) and \(\alpha = 0.694\); (\textit{a3}) Left part with \(C = 1.114 \times 10^{-3}\) and \(\alpha = 0.392\), right part with \(C = 8.189 \times 10^{-3}\) and \(\alpha = 0.677\); (\textit{a5}) \(C = 2.139 \times 10^{-4}\) and \(\alpha = 2.645\). The parameters of exponential decay Eq.~(\ref{ED}):  (\textit{a4}) Left part with \(D = 1.463 \times 10^{-3}\) and \(\beta = 0.577\), right part with \(D = 2.222 \times 10^{-3}\) and \(\beta = 7.689\); (\textit{a6}) \(D = 8.192 \times 10^{-4}\) and \(\beta = 1.038\). In all panels, the solid blue line represents the asymptotic $S_k$, where the dashed red line denote the fitting result.  }
\label{Fit3}
\end{figure}

Next, we will conduct a detailed examination of asymptotic wave-action spectrum $S_k$ in each region. To begin with, we adopt the following functional form to model the power-law decay region:
\bee\label{AD}
S_k=C|k+3|^{-\alpha}.
\ene
As for the exponentially decaying region, we employ the following functional representation:
\bee\label{ED}
S_k=De^{-\beta|k+3|}.
\ene
It is worth noting that the parameters \(\{C, D, \alpha, \beta\}\) vary across different regions. However, for simplicity, we will refrain from introducing separate parameters for each region:
 \begin{itemize}
 \vspace{0.05in}  \item The first region is near \(k=-3\), spanning \([-3.5, -2.5]\), and exhibits power-law decay, as shown in figure.~\ref{Fit3}(\textit{a1,a2}). In figure~\ref{Fit3}(\textit{a1}), we consider a global fit and obtain the parameters \(C = 9.061 \times 10^{-4}\) and \(\alpha = 0.667\). However, we find that splitting the fit on either side of \(k=-3\) yields better results (see figure~\ref{Fit3}(\textit{a2})). On the left side, the parameters are \(C = 9.856 \times 10^{-4}\) and \(\alpha = 0.638\), while on the right side, the corresponding parameters are \(C = 8.332 \times 10^{-4}\) and \(\alpha = 0.694\). We believe that this anisotropy can be mitigated by improving the computational environment (e.g., larger ensemble sizes, extended spatial domains).

   \vspace{0.05in} \item The first part of the second region is \(k \in [-5.1, -3.5]\), shown in the left part of figure~\ref{Fit3}(\textit{a3}), where the fitting parameters are \(C = 1.114 \times 10^{-3}\) and \(\alpha = 0.392\).  The second part of the second region is \(k \in [-2.5, -1.2]\), displayed in the right part of figure~\ref{Fit3}(\textit{a3}), with fitting parameters \(C = 8.189 \times 10^{-4}\) and \(\alpha = 0.677\).
  In fact, the second region  also appear in the NLS case~\cite{Zak15,Zak16}. However, in that context, the wave-action spectrum across these segments is uniform, while here it exhibits anisotropy. This discrepancy arises primarily because the part \(k \in [-2.5, -1.2]\) lies closer to \(k = 0\). Since \(k = 0\) remains stable in both the linear and nonlinear stages of MI, this segment exhibits a steeper decay. By contrast, in the NLS case, such anisotropy does not occur because the plane wave they consider is already centered at \(k = 0\).

   \vspace{0.05in} \item In the third region, \(S_k\) exhibits exponential decay, as shown in figure~\ref{Fit3}(\textit{a4}). The first segment, spanning \(k \in [-20, -5.1]\), is characterized by the fitted parameters \(D = 1.463 \times 10^{-3}\) and \(\beta = 0.577\). The second segment, \(k \in [-0.8, -0.1]\), yields the parameters \(D = 2.222 \times 10^{-3}\) and \(\beta = 7.689\), which represents a much faster decay compared to the first segment. This more rapid decay is primarily due to the proximity to the wave number \(k=0\), which remains stable throughout the entire process.

   \vspace{0.05in} \item The fourth region, corresponding to \(k \in [1.1, 2.6]\) (see figure~\ref{Fit3}(\textit{a5})), exhibits power-law decay, with fitted parameters \(C = 2.139 \times 10^{-4}\) and \(\alpha = 2.645\). This region has not been observed in the NLS case, and its emergence is primarily attributed to the wave number range of the linear MI described by Eq.~(\ref{Ra}).

   \vspace{0.05in} \item The fifth region, spanning \(k \in [2.6, 12]\) (see figure~\ref{Fit3}(\textit{a6})), demonstrates exponential decay, characterized by the parameters \(D = 8.192 \times 10^{-4}\) and \(\beta = 1.038\).
 \end{itemize}

 \vspace{0.05in} These results for the wave-action spectrum of the asymptotic steady state are summarized in Table \ref{TSD}.

\begin{table}
\renewcommand{\arraystretch}{1.5}
\centering
\begin{tabular}{|c|c|c|}
\hhline{|---|}
{\diagbox[width=12em,height=2.5em]{Wave number}{Parameters}} & $[C,\alpha]$ & $[D,\beta]$ \\
\hhline{|---|}
$[-3.5,\, -2.5]$ & $[9.061 \times 10^{-4},\, 0.667]$ & None \\
\hhline{|---|}
${[}-5.1,\, -3.5]$ & ${[}1.114 \times 10^{-3},\, 0.392]$ & None \\
\hhline{|---|}
${[}-2.5,\, -1.2]$ & ${[}8.189 \times 10^{-4},\, 0.677]$ & None \\
\hhline{|---|}
${[}-20,\, -5.1]$ & None & ${[}1.463 \times 10^{-3},\, 0.577]$ \\
\hhline{|---|}
${[}-0.8,\, -0.1]$ & None & ${[}2.222 \times 10^{-3},\, 7.689]$ \\
\hhline{|---|}
${[}1.1, \,2.6]$ & ${[}2.139 \times 10^{-4},\, 2.645]$ & None \\
\hhline{|---|}
${[}2.6, \, 12]$ & None & ${[}8.192 \times 10^{-4},\, 1.038]$ \\
\hhline{|---|}
\end{tabular}
\caption{The fitting parameters in different regions. The first column corresponds to the wave number range. The second column contains the parameters for power-law decay (see Eq.(\ref{AD})), while the third column lists the parameters for exponential decay (see Eq.(\ref{ED})), where none represents the wave-action spectrum does not correspond to this type of function.}
\label{TSD}
\end{table}

\begin{figure}
\centering
{\scalebox{0.5}[0.5]{\includegraphics{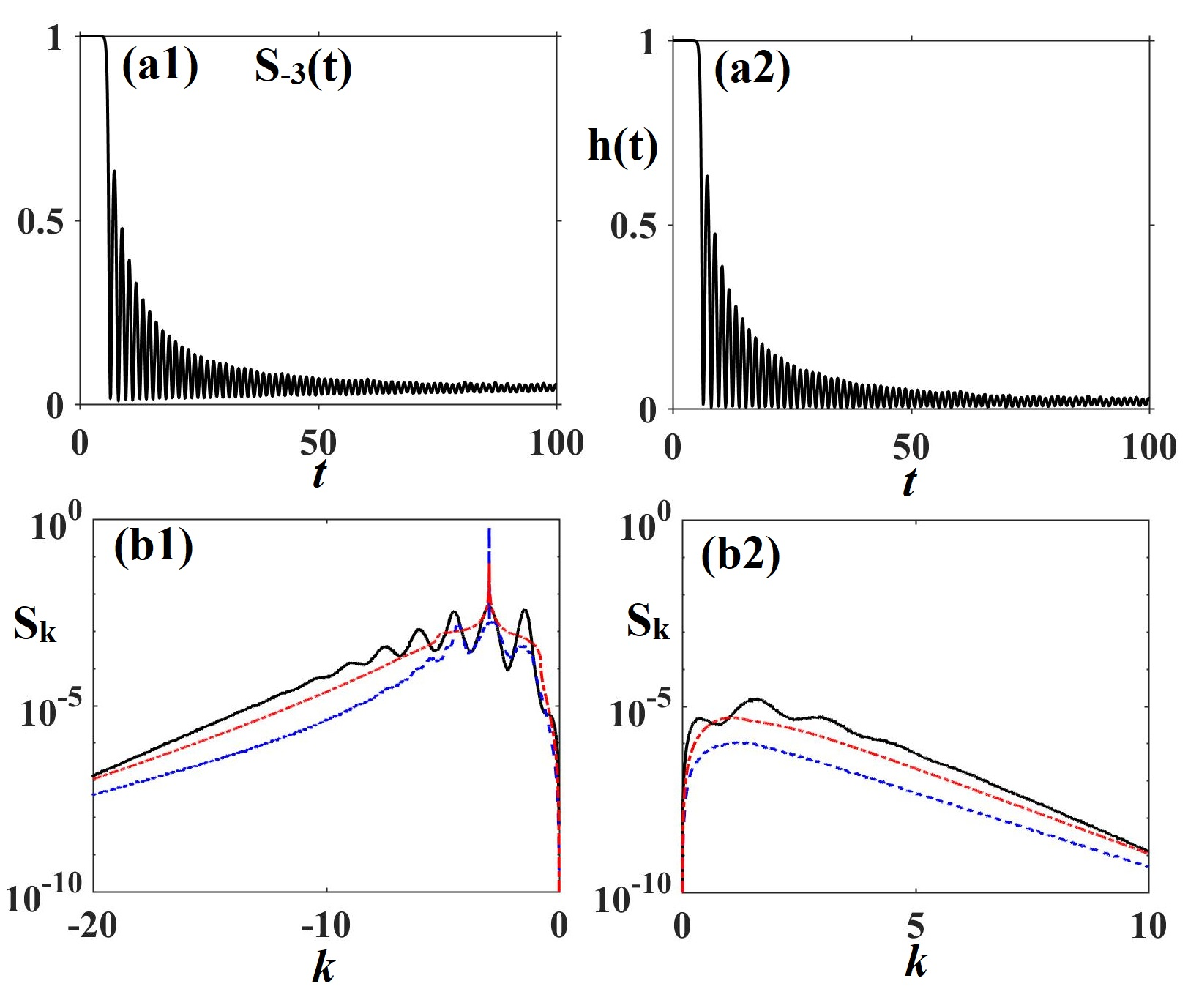}}}
\caption{ (\textit{a1}) The evolution of the wave-action spectrum at $k=-3$, i.e., $S_{-3}(t)$. (\textit{a2}) The evolution of the peak at $k=-3$ harmonic $h(t)$. (\textit{b1},\textit{b2}) The wave-action spectrum  function $S_k(t)$ is presented at three specific points: at \(t = 6.4\), corresponding to the first local maximum of the ensemble-averaged potential energy modulus $|\langle H_4\rangle|$ (solid black line); at \(t = 7.3\), corresponding to the first local minimum of $|\langle H_4\rangle|$ (dashed blue line); and as the asymptotic spatial correlation function (dash-dot red line). }
\label{Sksingle}
\end{figure}

During the linear phase of MI, and for an extended period in the nonlinear phase, the wave-action spectrum exhibits a discontinuity at \(k = -3\), manifested as a prominent peak that occupies only the $k=-3$ harmonic, as shown in figure~\ref{Sksingle}(\textit{a1}). This peak arises from the initial conditions Eq.~(\ref{IC}). Notably, while this peak does not vanish completely in the nonlinear phase, rather than it decays  in an oscillatory manner. We also measure the peak by calculating the difference between the  harmonic at \( k = -3 \) and the arithmetic average of the two neighboring harmonics:
\bee\label{Nei}
h(t)=S_{-3}(t)-\frac{1}{2}\left[S_{-3+ 2\pi / L}(t)+S_{-3-2 \pi / L}(t)\right] .
\ene
It can be observed that $h(t)$ also evolutions as  an oscillatory manner with approaching its asymptotic value from figure~\ref{Sksingle}(\textit{a2}).  At the same time, we observe that \( S_{-3}(t) \) and \( h(t) \) are in anti-phase with \( |\langle H_4 \rangle| \).

\begin{figure}
\centering
{\scalebox{0.45}[0.45]{\includegraphics{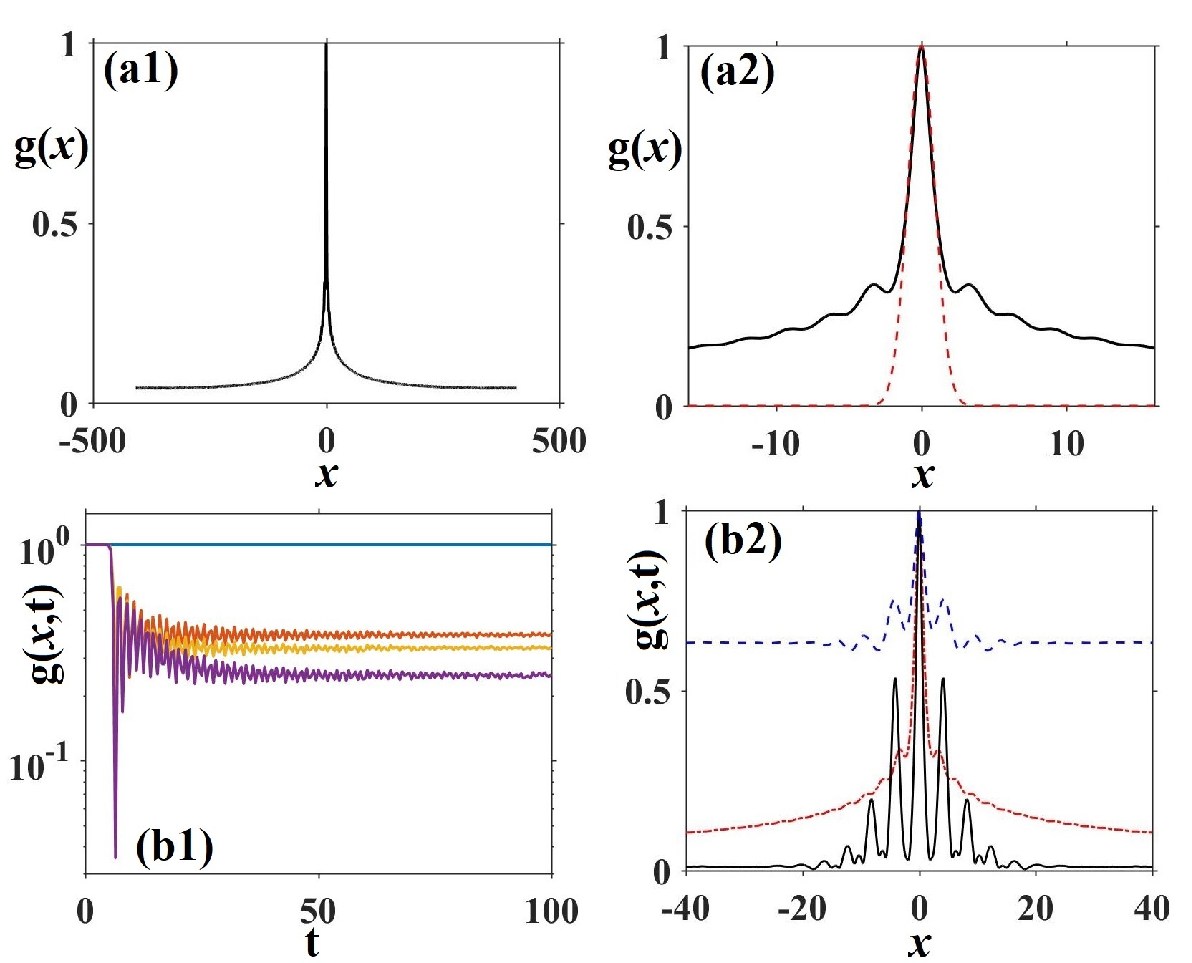}}}
\caption{ (\textit{a1}) The asymptotic auto-correlation function $g(x)$. (\textit{a2}) The asymptotic auto-correlation function $g(x)$ fitting by function $e^{-\frac{x}{2\sigma^2}}$  with $\sigma=0.935$. (\textit{b1}) Evolution of spatial correlation function $g(x,t)$ at $x = 0$
(blue), $x = \pi/2$ (red), $x=\pi$ (yellow), $x = 2\pi$ (purple).  (\textit{b2}) The spatial correlation function \(g(x, t)\) is presented at three specific points: at \(t = 6.4\), corresponding to the first local maximum of the ensemble-averaged potential energy modulus $|\langle H_4\rangle|$ (solid black line); at \(t = 7.3\), corresponding to the first local minimum of $|\langle H_4\rangle|$ (dashed blue line); and as the asymptotic spatial correlation function (dash-dot red line).}
\label{Auto}
\end{figure}

Next, we specifically examine \( S_k(t) \) at the moments corresponding to the strongest and weakest nonlinearity, which occur at \( t = 6.4 \) and \( t = 7.3 \), respectively, as depicted in figure~\ref{Sksingle}(\textit{b1,b2}). One can easily observe that at \( t = 6.4 \), \( S_k \) exhibits more peaks compared to one for \( t = 7.3 \).

A more advanced statistical measure related to the wave-action spectrum is the autocorrelation function, which can be expressed as~\cite{Zak15,Zak16,Sur19,Sur21,Sun21}
\begin{equation}\label{AutoF}
g(x, t) = \left\langle \frac{1}{L} \int_{-L/2}^{L/2} \psi^*(\eta-x, t)\psi(\eta, t) \, d\eta \right\rangle,
\end{equation}
which serves as a descriptor of turbulence correlations over spatial separations. This measure is instrumental in characterizing spatial coherence, fluctuation patterns, and the underlying dynamics of energy distribution, mixing, and structural development in turbulent systems. Furthermore, the autocorrelation function is directly linked to the wave-action spectrum through the inverse Fourier transform:
\bee\label{Re}
g(x, t) = \mathcal{F}_{D}^{-1}\left(S_k(t)\right).
\ene
The asymptotic auto-correlation function $g(x)$ is depicted in figure~\ref{Auto}(\textit{a1}). It is noted that \( g(x,t) = 1 \) at \( x = 0 \) for  all $t>0$,  primarily because
\bee
g(0,t)=\left\langle \frac{1}{L} \int_{-L/2}^{L/2} |\psi(\eta, t)|^2 \, d\eta \right\rangle=\left\langle N \right\rangle=1,
\ene
where the last equality comes from the fact that mass $N$ is a conserved quantity.
One notable feature of the spatial correlation function is that it remains at a distinct, nonzero value as \(|x| \to L/2\). This phenomenon is observed not only during the linear phase of the MI, but also persists for an extended duration into the nonlinear phase. Such behavior arises from the prominent peak at the $k=-3$ harmonic in the wave-action spectrum $S_k$. However, for \( |x| < 1.15 \), the asymptotic \( g(x) \) can be well approximated by the following Gaussian function:
\bee\label{FA}
g(x) \approx \exp\left(-\frac{x^2}{2\sigma^2}\right)
\ene
with $\sigma=0.935$ (see figure~\ref{Auto}(\textit{a2})).

For the given $x$, we display the time evolution of $g(x,t)$  depicted in figure~\ref{Auto}(\textit{b1}) for different $x$. Similarly, we observe that its evolution also oscillates, approaching to their asymptotic values. Although the oscillations here appear somewhat irregular, we have reason to believe that as the number of ensembles increases, the oscillation pattern will become inversely related to the absolute value of the ensemble-averaged potential energy, \( |\langle H_4 \rangle| \).  It should be noted that, during the linear MI stage (up to  $t\approx5$), we have
 \bee
 g(x,t)\approx1,
 \ene
  which is mainly due to $|\psi(x,t)|\approx1$. We then also examine the auto-correlation function $g(x)$ at the first local maximum and minimum of the ensemble-averaged potential energy modulus $|\langle H_4\rangle|$, as shown in figure~\ref{Auto}(\textit{b2}).  We observe that at \( t = 6.4 \), \( g(x) \) exhibits more peaks compared to \( t = 7.3 \), which is also determined by the characteristics of PDF (see figure~\ref{PDFDouble}(\textit{a1,a2})).

\section{Conclusions and discussions}
\label{CON}
In conclusion, our study of IT phenomena and RW emergence within the DNLSE provides valuable insights into their dynamics. We begin with a plane wave perturbed by a small noise, which led to an in-depth analysis of stationary turbulence using various statistical measures. Moreover, during the evolution of these statistical indices, they consistently oscillate and approach their asymptotic values. The asymptotic form can generally be expressed in a unified manner (see Eqs.~(\ref{Fit}) and (\ref{Fit2})).
Notably, we found that, unlike the symmetric turbulence observed in the NLSE, the DNLSE exhibits the asymmetric turbulence, where the energy spectrum shows different power-law decay in different regions. This is primarily due to the asymmetric between the wave number of the plane wave from the MI and the perturbation wave number.

As for potential directions for extending this research, several avenues warrant consideration. First, it would be valuable to explore whether other initial value excitations of IT could be considered, which might generate higher amplitude RWs, such as the coherent initial conditions proposed in work~\cite{Gel18}. Second, it would be interesting to investigate whether these phenomena similar to intermittency~\cite{Sur14}, often observed in fluid turbulence~\cite{WTbook1,WTbook2}, could occur in the IT of the DNLSE. Finally, to the best of our knowledge, there has been no systematic study of IT in higher-dimensional systems, such as the (2+1)-dimensional KP-1 equation.

\addcontentsline{toc}{section}{Appendix A}

\setcounter{equation}{0}
\renewcommand{\theequation}{A.\arabic{equation}}

	\vspace{0.2in}
\noindent {\bf Appendix A.\, Numerical Method for the DNLS Eq.~(\ref{DNLS})}
\label{NM}

In this work, we primarily employ the fourth-order split-step~\cite{Yang10} method for numerical experiments. Specifically, the DNLS Eq.~(\ref{DNLS}) can be split into the following linear and nonlinear parts:
\begin{equation}\label{Num}
\psi_t = (L + N)\psi
\end{equation}
where $L\psi = i\psi_{xx}$ and $N\psi = -\beta(|\psi|^2\psi)_x$.
Generally speaking, the split-step algorithm can be formulated as
\begin{equation}
\exp\left(dt(L+N)\right) \approx \prod_{k=1}^{n} \exp(\beta_k dt N) \exp(\alpha_k dt L),
\end{equation}
where $dt$ represents the time step size, and $\beta_k$, $\alpha_k$ ($k=1,2,\ldots,n$) denote the step coefficients for the nonlinear and linear operators, respectively. Here we adopt a fourth-order split-step algorithm, i.e., $n=4$ in the general formulation.

In this case, the split equations to be solved are
\begin{align}
v_t &= L(v) \label{eq:linear_split} \\
 \quad v_t &= N(v). \label{eq:nonlinear_split}
\end{align}
Here the split equation $v_t = L(v)$ can be solved by the discrete Fourier transform, and the split equation $v_t = N(v)$ has the exact solution formula
\begin{equation}
v(x,t) = v(x,0)\exp\left(-i\beta(2v_xv^*+vv_x^*) t\right)
\end{equation}
for the DNLS equation. During the numerical experiments, we monitor conserved quantities such as the $L^2$-norm Eq.~(\ref{LN}) and the Hamiltonian Eq.~(\ref{Ha}) to ensure the stability of the numerical algorithm.\\

\vspace{0.1in}
\noindent {\bf Funding}
 This work was partially supported by the National Natural Science Foundation of China (No. 12471242).

\vspace{0.1in}
\noindent {\bf Author Contributions} All authors contributed to the research and preparation of this work.

\vspace{0.1in}
\noindent {\bf Data Availability}  No new data were created or analysed in this work.

\vspace{0.1in}
\noindent {\bf  Declarations}

\noindent {\bf Conflict of interest} The authors declare no conflict of interest.

\end{document}